\documentclass[preprintnumbers, showpacs, floatfix,twocolumn,
preprintnumbers, superscriptaddress]{revtex4}
\usepackage{amsfonts}
\usepackage{amsmath}
\usepackage{amssymb,epsf}
\usepackage{latexsym}
\usepackage{graphicx,epsfig}
\usepackage{amssymb}
\usepackage{subfigure}
\usepackage[colorlinks=true,citecolor=blue,linkcolor=blue,urlcolor=black]{hyperref}
\usepackage{epstopdf}
\usepackage{color}

\def\be{\begin{equation}}
\def\ee{\end{equation}}
\def\ba{\begin{eqnarray}}
\def\ea{\end{eqnarray}}

\def\ra{\rangle}

\begin{document}
\title{A strong-weak coupling duality between two perturbed quantum many-body systems; CSS codes and Ising-like systems}
\author{Mohammad Hossein Zarei}
\email{mzarei92@shirazu.ac.ir}
\affiliation{Physics Department, College of Sciences, Shiraz University, Shiraz 71454, Iran}

\begin{abstract}

Graphs and recently hypergraphs have been known as an important
tool for considering different properties of quantum many-body
systems. In this paper, we study a mapping between an important class of
quantum systems namely quantum Calderbank-Shor-Steane (CSS) codes
and Ising-like systems by using hypergraphs. We show that the
Hamiltonian corresponding to a CSS code on a hypergraph $H$ which is perturbed by a uniform magnetic field is mapped to Hamiltonian of a
Ising-like system on dual hypergraph $\tilde{H}$ in a transverse
field. Interestingly, we show that a strong regime of couplings in
one of the systems is mapped to a weak regime of couplings in
another one. We also give some applications for such a mapping where we study
robustness of different topological CSS codes against a uniform
magnetic field including Kitaev's toric codes defined on graphs and color codes in
different dimensions. We show that a perturbed Kitaev's toric code on an arbitrary graph is mapped to an Ising model in a transverse
field on the same graph and a perturbed color code on a $D$-colex
is mapped to a Ising-like model on a $D$-simplicial lattice in a
transverse field. In particular, we use these results to
explicitly compare the robustness of TC on different graphs in
different dimensions. Interestingly, our results show that the
robustness of such topological codes defined on graphs decreases with increasing
dimension. Furthermore, we also use the duality mapping for some
self-dual models where we exactly derive the point of phase
transition.
\end{abstract}


\maketitle

\section{Introduction}
Graph theory has many applications in different fields of physics where graphs are usually used for illustrating two-body interactions \cite{graphbook}. Especially, in quantum information theory, graphs have been used to characterize some specific entangled states namely graph states \cite{Briegel2004}. Graph states are a set of stabilizer states which have especially been considered for quantum error correction where information are encoded in such states to be protected against decoherence \cite{gott1997}. Furthermore, recently an interesting extension of graphs called hypergraphs \cite{Berg} have attracted much attention because of their ability for illustrating many-body interactions in quantum many-body systems. For example, it has been shown that quantum entangled states can be encoded in the structure of a hypergraph \cite{hyp2013}. In particular, quantum hypergraph states have been introduced and many applications have been considered \cite{Brus2013}. Mapping quantum states to a hypergraphs helps one to use some relations in graph theory to study different properties of quantum entangled states \cite{hyp1, hyp2, hyp3, hyp4}.

On the other hand, since quantum entangled states can also been considered as the ground state of a quantum many-body system, most concepts in condense matter physics can also be used for quantum states. Among such quantum systems, quantum CSS codes introduced by Shor and et al \cite{Calderbank1996, Stean} have been attracted much attention. In particular, quantum CSS codes with topological order including Kitaev's toric codes (TC) and color codes (CC) play a key role because of natural robustness against local perturbations \cite{Kitaev2003, bombin2006}. Topological order is a new phase of matter that can not be characterized by symmetry breaking theory \cite{Wen1990}. Unlike the ordinary order, in a topological phase there is not a local order parameter and topological order should be characterized by some topological properties \cite{Wen1991, Hansson2004, Rokhsar1988}. Especially, degeneracy of the ground state is a measure of topological order that is related to topological structure of the physical system. While complete characterization of topological order is still an open problem, power of quantum codes with topological order for fault-tolerant quantum computation has been a popular field of research in recent years \cite{16, preskill2002, freedman2002, 17, 18, bravi2010}.

Studying robustness of TC and CC against local perturbations has been considered in many interesting recent works where a quantum phase transition occurs from a topological phase to a trivial phase. On the one hand different perturbations like uniform magnetic field and Ising perturbation have been studied \cite{vidal2011, jahromi2013, kargar2013, karimipour, zarei2015, zarei2016} and on the other hand topological states with three-level particles have been considered \cite{mohseni}. However, most topological CSS codes considered in the above researches are two-dimensional ones. Since, it has been shown that topological CSS codes in low dimensions do not have enough power for self-correcting quantum memory \cite{bravi}, different properties of topological codes in higher dimensions have been found more importance \cite{Bacon, bombin2016, review2016, zarei2017}.

In this paper, we use the idea of mapping quantum many-body systems to hypergraphs in order to consider a general quantum CSS code in presence of a uniform magnetic field. To this end, we begin with giving a formalism for mapping quantum CSS codes and Ising-like systems to hypergraphs where we use an approach introduced in \cite{zareim}. Then we consider a Hamiltonian as a weighted sum of stabilizers from the CSS code defined on a hypergraph $H$ in precense of a uniform magnetic field term. Next, we introduce a new basis to re-write the above Hamiltonian. We show that the above Hamiltonian in the new basis is equal to an Ising-like system with many-body interactions in a transverse field on dual hypergraph $\tilde{H}$. Interestingly, we show that such a mapping is a strong-weak coupling duality where the original model in a strong coupling regime is mapped to the new one in a weak coupling regime. In this way, we will give a duality relation that relates the problem of robustness of a CSS code on the $H$ against a strong magnetic field to the problem of a quantum phase transition in a Ising-like system in a transverse weak field on the $\tilde{H}$.

As an important application of our mapping, we will be able to
consider robustness of topological CSS codes in high dimensions
\cite{delgad, Bombin2015, Bravi2011, Sapta2009}. We show that the
problem of robustness of TC defined on an arbitrary graph in a uniform
magnetic field is mapped to a quantum phase transition in an Ising
model in a transverse field on the same graph. Furthermore, We
also give another example of our mapping for a CC on a D-colex
which is mapped to a Ising-like model with $(D+1)$-body
interaction on a D-simplicial lattice. Our explicit studies for TC
on different graphs, where we use well-known results on the
transverse Ising models \cite{trans}, show that robustness of TC defined on graphs
against a uniform magnetic field decreases in higher dimensions.
Although such a result is derived for a subclass of TC with qubits living on edges of a graph, it can lead to a new insight in application of
topological codes for quantum memory. Especially, it has been
shown \cite{bombin2016} that power of topological codes for
self-correcting quantum memory in finite temperature increases in
higher dimensions. Therefore, according to our results, increasing
dimension might not be necessarily good for a topological memory
where it improves the self-correctness of the code in finite
temperature but it might reduce the robustness against local
perturbation at zero temperature. 

Finally, similar to most duality relations, the above mapping is
useful for exactly determining the phase transition point
(robustness) of self-dual models. Corresponding to self-dual
hypergraphs, we introduce several self-dual models in different
dimensions where the phase transition occurs at critical ratio
$(\frac{h}{J})_c =1$. Especially, the one-dimensional case is the
well-known model of one-dimensional Ising model in a transverse
field.

The rest of the paper is as follows: In section (\ref{s1}) we give a brief review on definition of hypergraphs and especially dual of hypergraphs. Furthermore, we define an orthogonality for hypergraphs. In section (\ref{s2})  we give a formalism for mapping an arbitrary quantum CSS state and also an Ising-like system to hypergraphs. In section (\ref{s3}) , we give the main result of the paper where we map Hamiltonian of a quantum CSS state in presence of a uniform field on hypergraph $H$ to Hamiltonian of a Ising-like model in a transverse field on dual hypergraph $\tilde{H}$. In section (\ref{s4}) , we give some interesting applications of our mapping for TC defined on graphs and CC defined on color complexes in arbitrary dimensions and several self-dual models.

\section{hypergraphs and their dual}\label{s1}
\begin{figure}[t]
\centering
\includegraphics[width=8cm,height=3cm,angle=0]{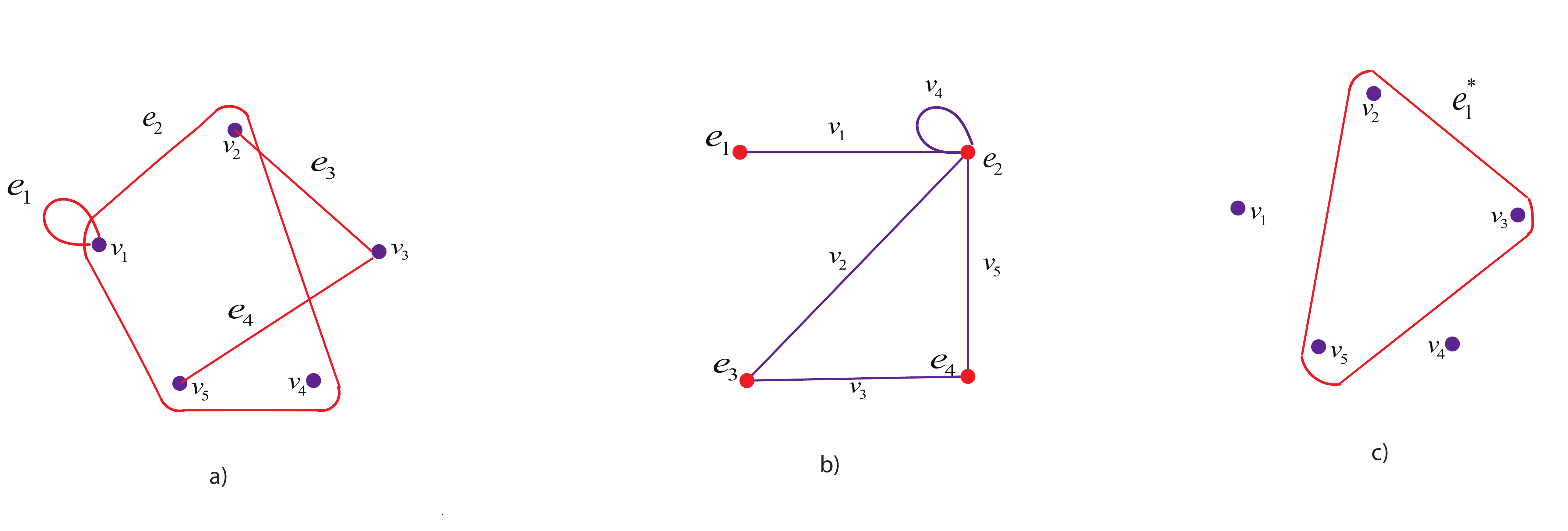}
\caption{(Color online) a) A simple example of a hypergraph where we denote vertices by black circles. we use a red loop for an edge containing only one vertex, a red link for edges containing two vertices and red closed curves for edges containing more than two vertices. b) Dual a hypergraph where edges play the role of vertices, denoted by red circles, and vertices play the role of edges, denoted by red curves, of dual hypergraph. c) Orthogonal hypergraph of the original hypergraph} \label{hyp}
\end{figure}

An ordinary graph $G=(V,E)$ is defined by two sets of vertices $V$ and edges $E$ where each edge connects two vertices of the graph. A hypergraph $H=(V,E)$, similar to a graph, is also defined by two sets of vertices and edges. However, unlike an ordinary graph, each edge of the hypergraph, which is called hyperedge, can involve arbitrary number of vertices. In other words, if degree of each hyperedge $e$ is equal to the number of vertices that are involved by $e$ and is denoted by $|e|$, the degree of edges of a hypergraph can be an arbitrary number. As an example in figure(\ref{hyp}-a), a hypergraph on five vertices $v_1, v_2 , v_3 , v_4 ,v_5$ has been shown where edge of $e_1 =\{v_1 \}$ is denoted by a loop , edge of $e_2 =\{v_1 , v_2 , v_4 , v_5\}$ is denoted by a closed curve and edges of $e_3 =\{v_2 , v_3\}$, $e_4 =\{v_3 , v_5 \}$ are denoted by links. In this figure, Degree of edges $e_1$, $e_2$, $e_3$, $e_4$ are equal to 1, 4, 2, 2, respectively.

For a hypergraph $H=(V,E)$ where $ V=\{v_1 ,v_2 , ...,v_K \} $ and  $E=\{e_1 , e_2 , ...,e_N \}$, there is a simple definition for a dual hypergraph $\tilde{H}$ . In fact, dual of a hypergraph $H(V,E)$ is a hypergraph $\tilde{H}=(\tilde{V},\tilde{E})$ where $\tilde{V}=\{ \tilde{v}_1 ,\tilde{v}_2 ,...\tilde{v}_N \}$ and $\tilde{E}=\{\tilde{e}_1 ,\tilde{e}_2 ,...,\tilde{e}_K \}$ where $\tilde{e}_i =\{ \tilde{v}_m ~|~ v_i ~\in ~e_m ~in~ H \}$. In other words, vertices and edges of $H$ are switched in the $\tilde{H}$. For example,  see figure(\ref{hyp}) for a hypergraph and its dual. A hypergraph is called self-dual if it and its dual are the same \cite{Berg}. It is clear that for a self-dual hypergraph, it is necessary that the number of vertices be the same as the number of hyperedges $K=N$.

It is also possible to introduce a binary vector representation
for hyperedges of a hypergraph. To this end, for a hypergraph with
$K$ vertices $v_1 , v_2 ,..., v_k$ and $N$ edges $e_1 , e_2 , ...,
e_N$, we  relate a binary vector, which is called edge vector, to
each hyperedge $e_m$. Such an edge vector will have $K$ components
which are denoted by $e_{m}^{j}$ and $j=\{1,2,3,...,K\}$ where
$e_{m}^{j}=1$ if $v_{j} \in e_m$ and $e_{m}^{j}=0$ otherwise. For
example, for hypergraph in figure(\ref{hyp}-a), binary vectors
corresponding to four edges are $e_1 =(1,0,0,0,0)$, $e_2
=(1,1,0,1,1)$, $e_3 =(0,1,1,0,0)$ and $e_4 =(0,0,1,0,1)$ . In this
way, we will have $N$ binary vectors corresponding to $N$ edges of
the hypergraph. Furthermore, in a binary representation, two edge
vectors can be added to achieve a new edge vector. If an edge
vector related to a hypergraph $H$ is equal to a superposition of
other edge vectors of the $H$, it will be called dependent. We
call a set of all independent edges of a hypergraph a independent
set which is denoted by $\mathcal{I}$. It is clear that
$|\mathcal{I}|\leq |V|$ which means that the number of independent
edges is always lesser that the number of vertices.

By binary vector representation of edges, For a hypergraph $H$,
one can define an orthogonal hypergraph $H^{*}$. To this end, two
hyperedges $e$ and $e'$ are called orthogonal if and only if their
corresponding binary vectors are orthogonal as $e.e'=0$. Now
consider a hypergraph $H=(V,E)$ with a independent set
$\mathcal{I}$ of edges.  We define orthogonal hypergraph
$H^{*}=(V^{*},E^{*})$ that has the same vertices of the $H$,
$V^{*}=V$, but it has $K-|\mathcal{I}|$ distinct edges that are
orthogonal to all edges of the $H$ where $|\mathcal{I}|$ is the
number of independent edges, see figure(\ref{hyp}-c).

 We can also show that, for any hypergraph $H$, there is certainly an orthogonal hypergraph $H^{*}$. To this end, consider binary vector corresponding to an edge $e^{*}$ of the hypergraph $H^{*}$. Since the $e^{*}$ should be orthogonal to all edges of the $H$, it is necessary to $e^{*}.e_{m}=0$ for $m=1,2,...N$. Since the number of independent edges is $|\mathcal{I}|$, it is enough to the above condition is held for $|\mathcal{I}|$ independent edges of the $H$. Therefore, these conditions are equal to a set of $|\mathcal{I}|$ independent equations on $K$ binary variables and it is clear that there are $K-|\mathcal{I}|$ independent answers for such a set of equations. These $K-|\mathcal{I}|$ answers are the edges of hypergraph $H^{*}$.

\section{Quantum many-body systems on hypergraphs; Ising-like systems and quantum CSS codes}\label{s2}
For a physical system with two-body interactions, graphs are a useful tool where each edge of the graph illustrates a two-body interactions between physical variables living in two end-points of that edge. Ising models with two-body interactions between Ising variables $S_i =\pm 1$ are an important example of application of graphs. However, there are also several important physical systems with many-body interactions. For example, one can consider Ising-like systems with many-body interactions between Ising variables. In order to encode interaction patterns of such systems, hypergraphs are a better candidate. Here, we relate a hyperedge to each interaction term of such a model. In this way, corresponding to a hypergraph $H=(V,E)$ we define a Ising-like system with the following Hamiltonian:
\begin{equation}
\mathcal{H}_I=\sum_{e\in E}\prod_{v\in e}X_i
\end{equation}
where qubits live in vertices of the $H$, $X$ is a pauli operator with eigenvalues $\pm 1$,  $e\in E$ refers to all hyperedges of the $H$ and $v\in e$ refers to all vertices belonging to the $e$.

Another example of a quantum many-body systems, that we consider in this paper, are quantum CSS codes. They are one of the most popular codes in quantum error correction that have been introduced by Shor and et al \cite{Calderbank1996, Stean}. A quantum CSS state on $K$ qubits is a stabilizer state that stabilized by $X$-type and $Z$-type operators belonging to Pauli group on $K$ qubits. Here we use hypergraphs to encode the structure of such states. To this end, consider a hypergraph $H=(V,E)$ and suppose that all hyperedges of the $H$ are independent and therefore $\mathcal{I}=E$ where $|V|=K$, $|E|=N$ and $K\geq N$. We insert $K$ qubits in all vertices of the hypergraph. Then we define a $X$-type operator corresponding to each independent edge of the hypergraph. we denote such operator by $A_e$ where, for each $e\in \mathcal{I}$, it is defined in the following form:
\begin{equation}
A_e =\prod_{i\in e}X_i
\end{equation}
where $i \in e$ refers to all vertices belonging to edge $e$. Now, we define the following state as a quantum CSS state corresponding to the $H$:
\begin{equation}\label{q1}
|C_H\ra=\frac{1}{2^{\frac{|\mathcal{I}|}{2}}}\prod_{e \in \mathcal{I}}(1+A_e )|0\ra ^{\otimes N}
\end{equation}
where $e\in \mathcal{I}$ refers to all independent edges belonging to the $H$ and $|0\ra$ is the positive eigenstate of the Pauli operator $Z$ where $Z|0\ra=|0\ra$. In order to prove that the above state is a CSS state, we should show that it is stabilized by $Z$-type and $X$-type operators. By the fact that $A_e (1+A_e)=(1+A_e)$, we conclude that all $A_e$ operators are stabilizers of the above state. In this way, since the number of independent edges of the hypergraph is equal to $N$, we have found $N$ number of stabilizers of the above state where all of them are $X$-type operators.

Since $K\geq N$, we should find $K-N$ numbers of other stabilizers
in order to completely characterize the CSS state. To this end
consider the orthogonal hypergraph of the $H$ that is
called $H^{*} $. We define $K-N$ number of $Z$-type operators
corresponding to each edge of the $H^{*}$. We denote them by
$B_{e^{*}}$ and define in the following form:
\begin{equation}
B_{e^{*}} =\prod_{i\in e^{*}}Z_i
\end{equation}

As we showed in section(\ref{s1}), all edges of the $H^{*}$ are orthogonal to the edges of the $H$. It is clear that when two edges are orthogonal to each other it is necessary that the number of common vertices of those edges is an even number. By this fact, it is simple to show that $[A_e , B_{e^{*}}]=0$. In this way, we have found all $K$ stabilizers corresponding to the state (\ref{q1}). Since all these operators are $X$-type or $Z$-type operators, it is concluded that the state (\ref{q1}) is a quantum CSS state.

There is also possible to represent the CSS state (\ref{q1}) by $Z$-type operators $B_{e{*}}$ in the following form:
\begin{equation}\label{q2}
|C_H\ra=\prod_{e^{*} \in E^{*}}(1+B_{e^{*}} )|+\ra ^{\otimes K}
\end{equation}
where $|+\ra$ is the positive eigenstate of the Pauli operator $X$. In this way, we can give a CSS state corresponding to each hypergraph with independent edges. Finally, such a state is the ground state of a Hamiltonian in the following form:
\begin{equation}\label{cssh}
\mathcal{H}_C=-J \sum_{e\in E}A_e -J \sum_{e^{*}\in E^{*}}B_{e^{*}}
\end{equation}

\section{Mapping quantum CSS codes to Ising-like systems; a strong-weak coupling duality}\label{s3}
 Duality in many-body systems is an interesting and well-stablished problem that has attracted much attention during past decades such as duality for generalized Ising models\cite{generalized} and recently duality in quantum models \cite{gaug1, gaug2}. In this section, we find a duality mapping from quantum CSS code in a uniform magnetic field to Ising-like systems in a transverse field. To this end, we begin with studying a CSS code defined on a hypergraph $H$ in presence of a uniform magnetic field. We consider a Hamiltonian for a CSS code where ground state of that Hamiltonian is stabilizer state of the quantum CSS code. Then we add a magnetic term to this Hamiltonian in the following form:
\begin{equation}\label{init}
\mathcal{H}=-J \sum_{e\in E}A_e -J \sum_{e^{*}\in E^{*}}B_{e^{*}} -\sum_{v\in V} h Z_v
\end{equation}
where $v \in V$ refers to vertices of the hypergraph. it is clear that when $\frac{h}{J}$ is equal to zero the ground state of the model is a quantum CSS state and when $\frac{h}{J}$ goes to infinity the ground state will be a product state in the form of $|00...0\ra$ where $|0\ra$ is eigenstate of Pauli operator $Z$. Now, we should study the problem of a quantum phase transition that can happen by tuning the magnetic field $h$ when there are two different phases in two limits $h=0$ and $h\rightarrow \infty$. We should emphasize that, in view of application in quantum information theory, such a phase transition is also related to the robustness of a quantum CSS code against a uniform magnetic field where the phase transition point is a measure of the robustness of the CSS code. Here, we give a general mapping for the above Hamiltonian to convert it to an Ising-like system in a switched regime of couplings. To this end, we use a new basis for re-writing the above Hamiltonian where we consider a new basis in the following form:
\begin{equation}\label{ab}
|C_{r_{1}, r_{2}, ..., r_{N}}\ra=\prod_{e \in E}(1+(-1)^{r_{e}}A_e )|0\ra ^{\otimes K}
\end{equation}
where $r_e$ are binary numbers that are related to each edge of the hypergraph and $N$ is the number of edges of the hypergraph. Since $K\geq N$ it is clear that the above basis is not a complete basis. In fact, For a complete basis we should also consider $Z$-type operators $B_{e^{*}}$ in the form of $\prod_{e^{*} \in E^{*}}(1+(-1)^{r_{e^{*}}}B_{e^{*}} )$. However, since the magnetic terms in the original Hamiltonian commute with operators $B_{e^{*}}$, we can stay in a subspace that is stabilized by all operators $B_{e^{*}}$. In this way we consider the bases (\ref{ab}) in such a subspace.

Finally, we are ready to re-write the original Hamiltonian in the new basis. To this end, we define vertices corresponding to binary variables $r_e$ related to each hyperedge of the $H$. In the other words, these new vertices denoted by $\tilde{v}$ are vertices of dual hypergraph $\tilde{H}$. In this way, the bases defined in (\ref{ab}) are equal to a computational basis on qubits which live in $\tilde{v}$ and we call them dual qubits.

 Then, we consider the effect of each term of the original Hamiltonian on the basis (\ref{ab}). Since the operators $B_{e^{*}}$ commute with $A_e$, we will have $B_{e^{*}}(1+(-1)^{r_e}A_e)=(1+(-1)^{r_e}A_e)B_{e^{*}}$. Then by the fact that $B_{e^{*}}|00..0\ra=|00..0\ra$, it is concluded that $B_{e^{*}}|C_{r_{1}, r_{2}, ..., r_{N}}\ra=|C_{r_{1}, r_{2}, ..., r_{N}}\ra$. In the next step, we consider the effect of operators $A_{e}$. Since $A_{e}^{2}=1$, we will have $A_{e}(1+(-1)^{r_e}A_{e})=(-1)^{r_e}(1+(-1)^{r_e}A_e)$. Therefore, the effect of $A_e$ in the new basis is similar to Pauli operator $Z$ on dual qubits. In this way, we can re-write term $A_e$ in the original Hamiltonian with $Z_e$ in the new basis on dual qubits.

The most interesting part of changing basis is related to magnetic terms in the original Hamiltonian. It is clear that $Z_v$ does not commute with those operators $A_e$ including vertex $v$. Since, In the dual space, vertex $v$ is a hyperedge of $\tilde{H}$ denoted by $\tilde{e}$ and edge $e$ is a vertex of $\tilde{H}$ denoted by $\tilde{v}$, we can denote operators $Z_v$ and $A_e$ by $Z_{\tilde{e}}$ and $A_{\tilde{v}}$. In this way, we can say that $Z_{\tilde{e}}$ does not commute with those operators $A_{\tilde{v}}$ that $\tilde{v}$ is a member of the hyperedge $\tilde{e}$ in the dual space. On the other hand if operator $Z_{\tilde{e}}$ does not commute with an operator $A_{\tilde{v}}$, we will have $Z_{\tilde{e}} (1+(-1)^{r_{\tilde{v}} }A_{\tilde{v}})=(1+(-1)^{r_{\tilde{v}} +1} A_{\tilde{v}})Z_{\tilde{e}}$. In this way the effect of operator $Z_{\tilde{e}}$ leads to rise the value of $r_{\tilde{v}}$ for all $A_{\tilde{v}}$ which does not commute with $Z_{\tilde{e}}$. It is equal to applying Pauli operators $X$ on all dual qubits $\tilde{v}$ belonging to the edge $\tilde{e}$. Therefore the magnetic term of the original Hamiltonian should be re-written in the new basis in the form of $h\sum_{\tilde{e}} \prod_{\tilde{v} \in \tilde{e}}X_{\tilde{v}}$. In this way the original Hamiltonian in the new basis will be in the following form:
\begin{equation}\label{main}
\mathcal{H}=-h\sum_{\tilde{e}} \prod_{\tilde{v} \in \tilde{e}}X_{\tilde{v}} - J\sum_{\tilde{v}}Z_{\tilde{v}}-K
\end{equation}

The first term in the above Hamiltonian is a Ising-like system on the $\tilde{H}$ with many-body interactions corresponding to each hyperedge of the $\tilde{H}$ and the second term is a transverse field. Interesting point is that the role of the magnetic field $h$ and coupling constant $J$ of the original Hamiltonian (\ref{init}) have been switched in the new Hamiltonian (\ref{main}). In this way, a regime of the strong magnetic field in the original model, where the ratio $\frac{h}{J}$ is bigger than $1$, is mapped to a regime of weak magnetic field in the dual Hamiltonian where the ratio $\frac{J}{h}$ is smaller than $1$. To this reason, we call our mapping a strong-weak coupling duality.

\section{Applications of the duality mapping}\label{s4}
In this section we give some applications for the above mapping. Especially, we use the duality mapping for studying the robustness of two important sets of CSS codes namely TC and CC. Finally, we introduce some self-dual models where we exactly derive the phase transition point.
\subsection{Robustness of TC against magnetic field}
\begin{figure}[t]
\centering
\includegraphics[width=8cm,height=3.5cm,angle=0]{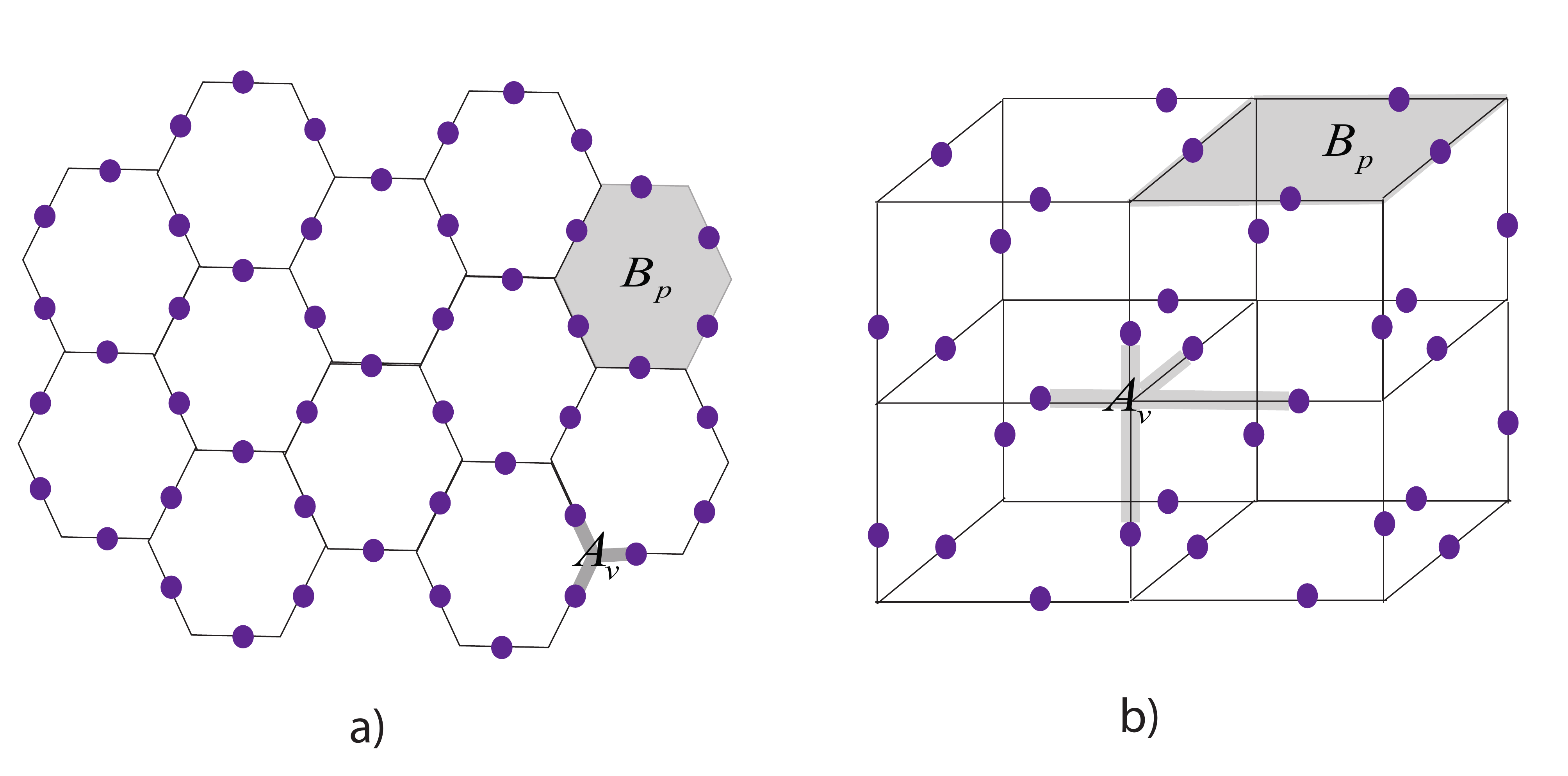}
\caption{(Color online) Two sample graphs where TC are defined. An $X$-type stabilizer is defined corresponding to each vertex of the graph and a $Z$-type stabilizer is defined corresponding to each plaquette of the graph} \label{kitaev}
\end{figure}

Here we consider TC which are defined on graphs where qubits live in edges of the graph. We emphasize that while TC can generally be defined on lattices with qubits living on higher dimensional cells, our results in this section are only related to a subclass of TC defined on graphs with edge qubits. To this end, corresponding to a graph $G$, two set of $X$-type and $Z$-type stabilizer operators are defined, see Fig(\ref{kitaev}) for two sample graphs, in the following form:
\begin{equation}
A_v =\prod_{i \in v} X_i ~~,~~B_p =\prod_{i\in \partial p}Z_i
\end{equation}
where $A_v$ corresponds to vertex $v$ of the graph and $i\in v$ refers to all qubits incoming to vertex $v$. Furthermore, $B_p$ corresponds to plaquette $p$ of the graph and $i\in \partial p$ refers to all qubits on boundary of plaquette $p$.
By the fact that $A_v (1+A_v)=(1+A_v)$ and $[B_p , A_v ]=0$, it is simple to check that the following state will be stabilized by the above operators:
\begin{equation}
|K\ra =\prod_{v\in \mathcal{I}}(1+A_v)|00...0\ra
\end{equation}
where we ignore the normalization factor and $\mathcal{I}$ refers to an independent set of $X$-type operators. It is well-known that such a state on a specific topological structure like a torus shows a topological degeneracy.

\begin{figure}[t]
\centering
\includegraphics[width=6cm,height=2cm,angle=0]{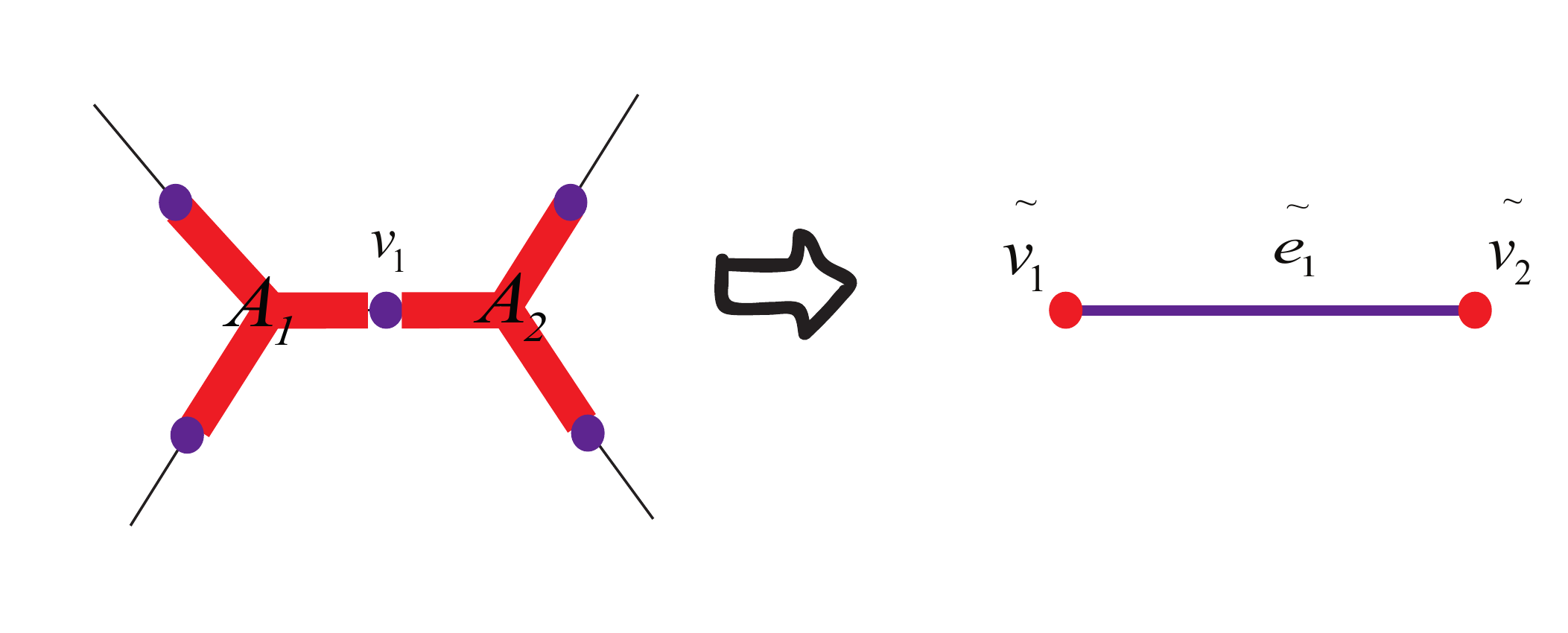}
\caption{(Color online) In a dual space we insert a red vertex corresponding to each hyperedge of the $H$. Since each vertex of a hexagonal lattice is a member of two neighboring hyperedges of the $H$, in the dual space a hyperedge involves two vertex of the $\tilde{H}$. Therefore, the $\tilde{H}$ is a ordinary graph which is matched on the original graph.} \label{kitexa}
\end{figure}

Now, let us consider such model in precense of a uniform magnetic field. Since TC is a CSS state, we should be able to use the dual mapping that we introduced in the previous section. To this end, we should give a hypergraph representation for TC. Therefore, we define a hypergraph $H$ with vertices corresponding to qubits of TC and edges corresponding to $X$-type operators $A_v$. In this way, corresponding to each vertex $v$ of the $G$, there is a hyperedge of the $H$ which involve all qubits belonging to the $v$, see Fig(\ref{kitexa}).

\begin{figure}[t]
\centering
\includegraphics[width=8cm,height=4.5cm,angle=0]{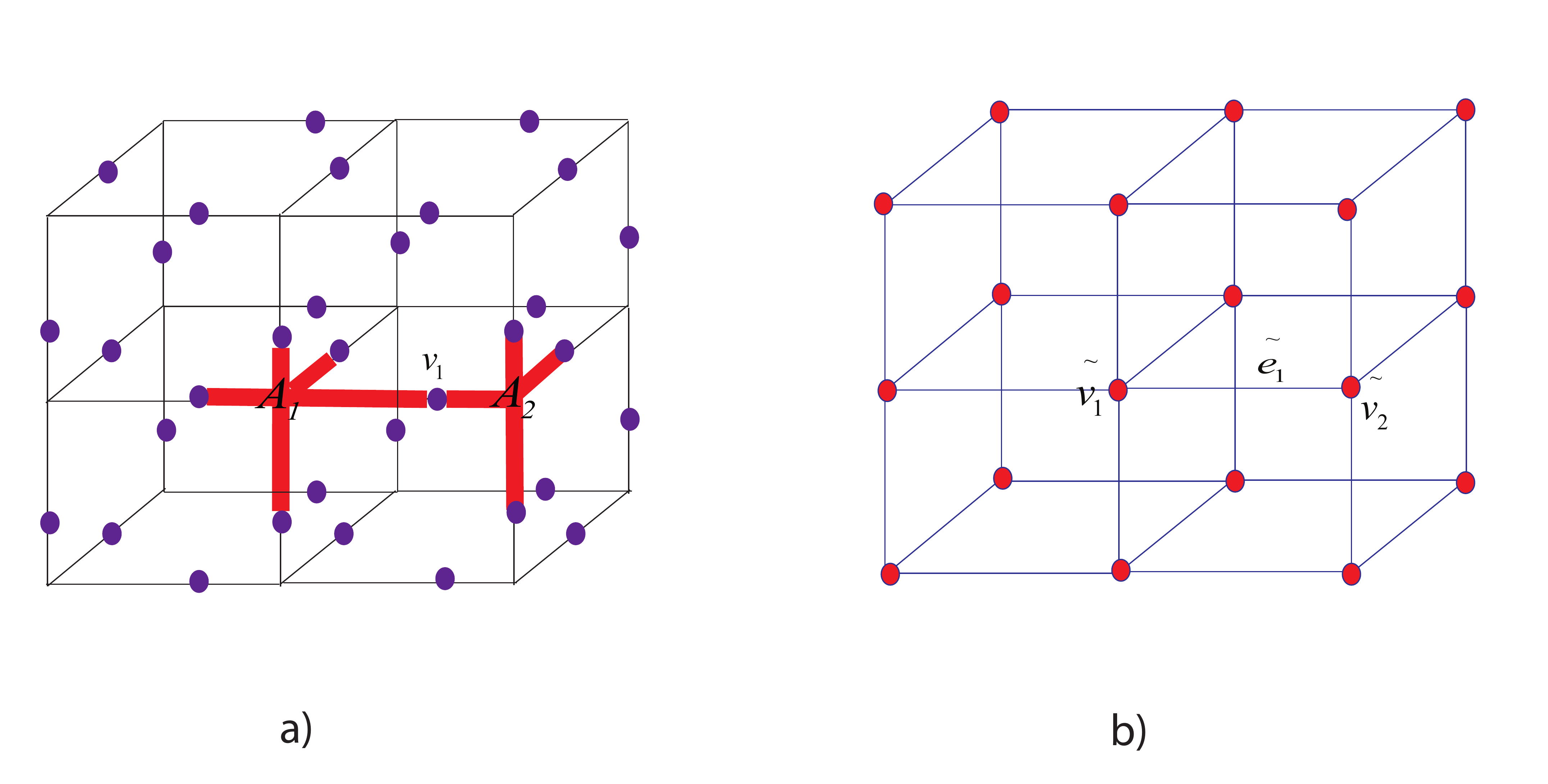}
\caption{(Color online) Similar to the 2D model in Fig.(\ref{kitexa}), In a dual space we insert a red vertex corresponding to each hyperedge of the $H$. Since each vertex of a 3D lattice is a member of two neighboring hyperedges of the $H$, in the dual space a hyperedge involves two vertex of the $\tilde{H}$. Therefore, the $\tilde{H}$ is a ordinary graph which is matched on the original 3D rectangular lattice.} \label{kit}
\end{figure}

In the next step, according to the dual mapping, the TC on the hypergraph $H$, corresponding to initial graph $G$, in a uniform magnetic field is mapped to a Ising-like system on dual hypergraph $\tilde{H}$ in a transverse field. Therefore, we should find the $\tilde{H}$ for the TC. To this end, we should find all hyperedges of the $H$ that a vertex of the $H$ is a member of them. Since each qubit of the TC lives in an edge of the original graph $G$, it is just included by two vertex operators corresponding to two neighboring vertices of the $G$, see Fig(\ref{kitexa}). Therefore, it is concluded that each vertex of the $H$ is a member of two neighboring hyperedges of the $H$. In a dual space, it means that each edge of the $\tilde{H}$ involves two vertices of the $\tilde{H}$. Therefore, it is enough to insert a vertex of the $\tilde{H}$ called $\tilde{v}$ in each vertex of the $G$ and then each edge of the $\tilde{H}$ involves two neighboring vertices on the $G$, see Fig(\ref{kitexa}). It means that dual hypergraph $\tilde{H}$ is a ordinary graph that is exactly matched on graph $G$. A spin model corresponding to such a hypergraph is Ising model in a transverse field, see also Fig(\ref{kit}) for a 3D example.

Finally we conclude that TC on an arbitrary graph in a magnetic
field is equal to Ising model on the same graph in a transverse
field. Interesting point is that Ising model in a transverse field
is a well-known model that has been studied in statistical
physics. Importantly, according to our mapping, the point of the
phase transition of Ising model on different lattices will
determine the robustness of the TC on the same graph against
magnetic field. We should also be care that while the robustness
of TC is determined by the ratio $\frac{h}{J}$, the quantum phase
transition point of Ising model in accord of the relation
(\ref{main}) is determined by $\frac{J}{h}$. In other words, in
the corresponding Ising model $J$ is power of transverse magnetic
field. By this point and well-known results derived in
\cite{trans}, we provide a table (\ref{tab}) for robustness of TC
on different lattices against a uniform magnetic field.
\begin{table}[t]
\centering
\begin{tabular}{ |p{4cm}|p{5cm}| }
\hline
 TC on different lattices & Robustness against a magnetic field\\
\hline

 On a honeycomb lattice & $(\frac{h}{J})_c \approx ~ 0.469$\\

\hline
On a kagome lattice &$(\frac{h}{J})_c \approx ~ 0.339$\\

\hline
 On a triangular lattice &  $(\frac{h}{J})_c \approx ~ 0.209$\\

\hline
On a square lattice & $(\frac{h}{J})_c \approx ~ 0.328$\\

\hline
 On a cubic lattice & $(\frac{h}{J})_c \approx ~ 0.194$\\

\hline
\end{tabular}
\caption{The robustness of TC on different lattices against a
uniform magnetic field.} \label{tab}
\end{table}

We should emphasize in an important result according to our
mapping. It is well-known that the phase transition point of
transverse Ising model increases in higher dimensions and it means
that the ratio $\frac{J}{h}$ increases in higher dimensions. Since
the robustness of TC is determined by the ratio $\frac{h}{J}$, we
conclude that the robustness of TC, defined on a graph with qubits living on edges, against a uniform magnetic
field decreases in higher dimensions. We believe that it is an
important result because, according to recent researches, power of
topological codes for a self-correcting quantum memory increases
in higher dimensions. Although it is clear that the measure of
self-correctness of a topological code in finite temperature is
different with the measure of the robustness at zero temperature,
our result proposes that increasing dimension might not be
necessarily a good strategy for improving a topological code. In
other words, by increasing dimension, on the one hand power of
self-correctness increases and on the other hand the robustness
decreases. Therefore there might be an optimum dimension for an
efficient topological memory.

\subsection{Robustness of CC against magnetic field}
\begin{figure}[t]
\centering
\includegraphics[width=8cm,height=4.5cm,angle=0]{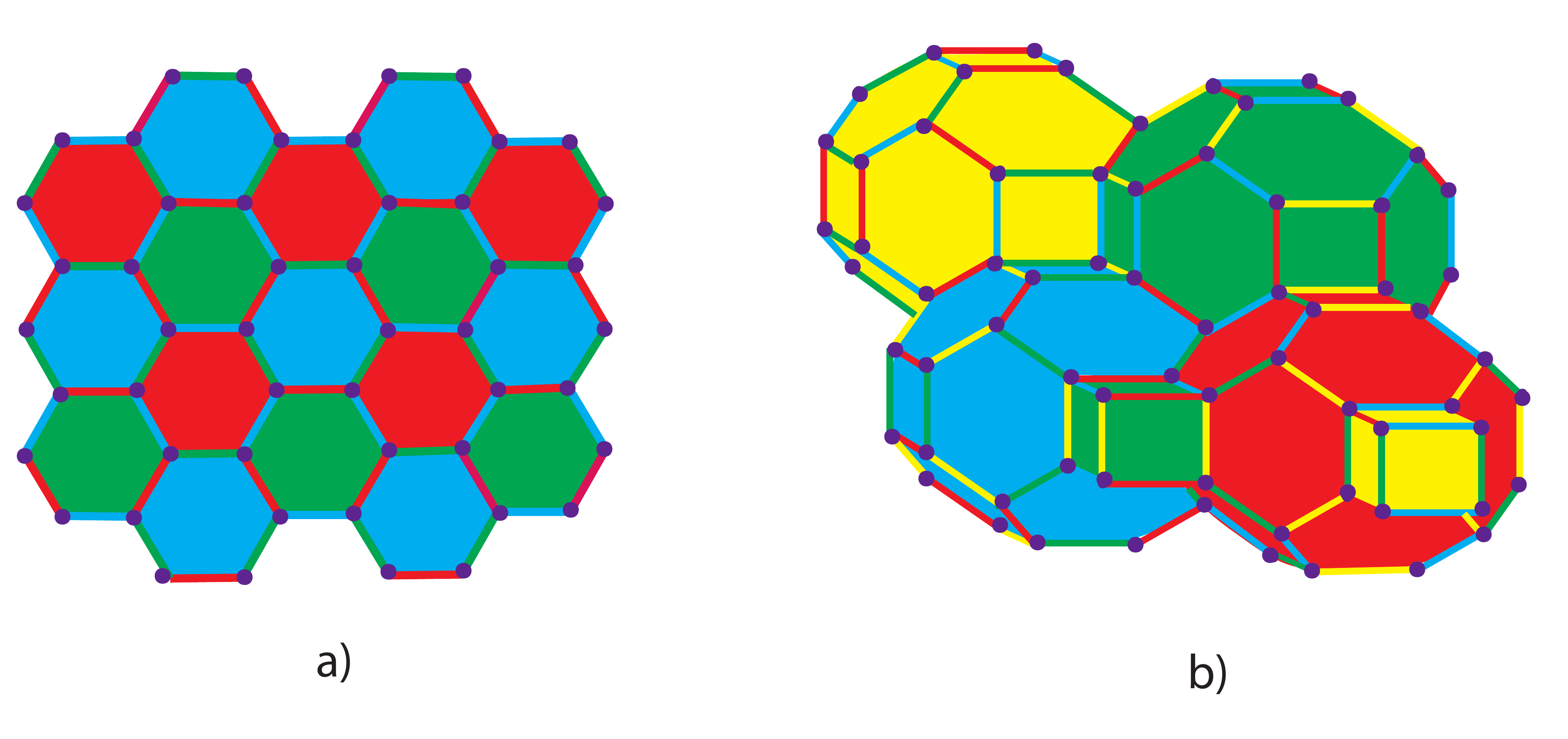}
\caption{(Color online) Two sample lattice where a color code can be defined. The left hand is a 2-colex (three-colorable hexagonal lattice) and the right-hand is a 3-colex with four-colorable cells. } \label{color}
\end{figure}

As another example of the dual mapping, here we consider robustness of CC against a uniform magnetic field. CC are defined on D-colexes that are color complexes on D-dimensional manifold where cells of the lattice can be colored by $D+1$ distinct colors \cite{delgad}. In Fig(\ref{color}), we show two examples of colexes in two and three dimension. A CC on a $D$-colex are defined by two set of $X$-type and $Z$-type stabilizers in the following form:
\begin{equation}
A_c =\prod_{i\in C}X_i ~~,~~B_{c'} =\prod_{i\in C'} Z_i
\end{equation}
where $C$ and $C'$ are cells of the colex so that $[A_c
,B_{c'}]=0$. For example for a 2-colex (a hexagonal lattice) the
above operators are defined corresponding to each plaquette of the
lattice. On the other hand, for a 3-colex $X$-type operators are
defined corresponding to each three-dimensional cell of the
lattice and $Z$-type operators are defined corresponding to each
plaquette of the lattice.

In order to consider the CC in a uniform magnetic field, we should represent the CC as a CSS code on a hypergraph. We explain this idea with two simple examples and then we extend it to more general cases. Therefore, consider a CC on a 2-colex like an hexagonal lattice. The CC corresponding to this lattice is in the following form:
\begin{equation}
|CC_2 \ra =\prod_{p}(1+A_p)|00...0\ra
\end{equation}

\begin{figure}[t]
\centering
\includegraphics[width=6cm,height=2cm,angle=0]{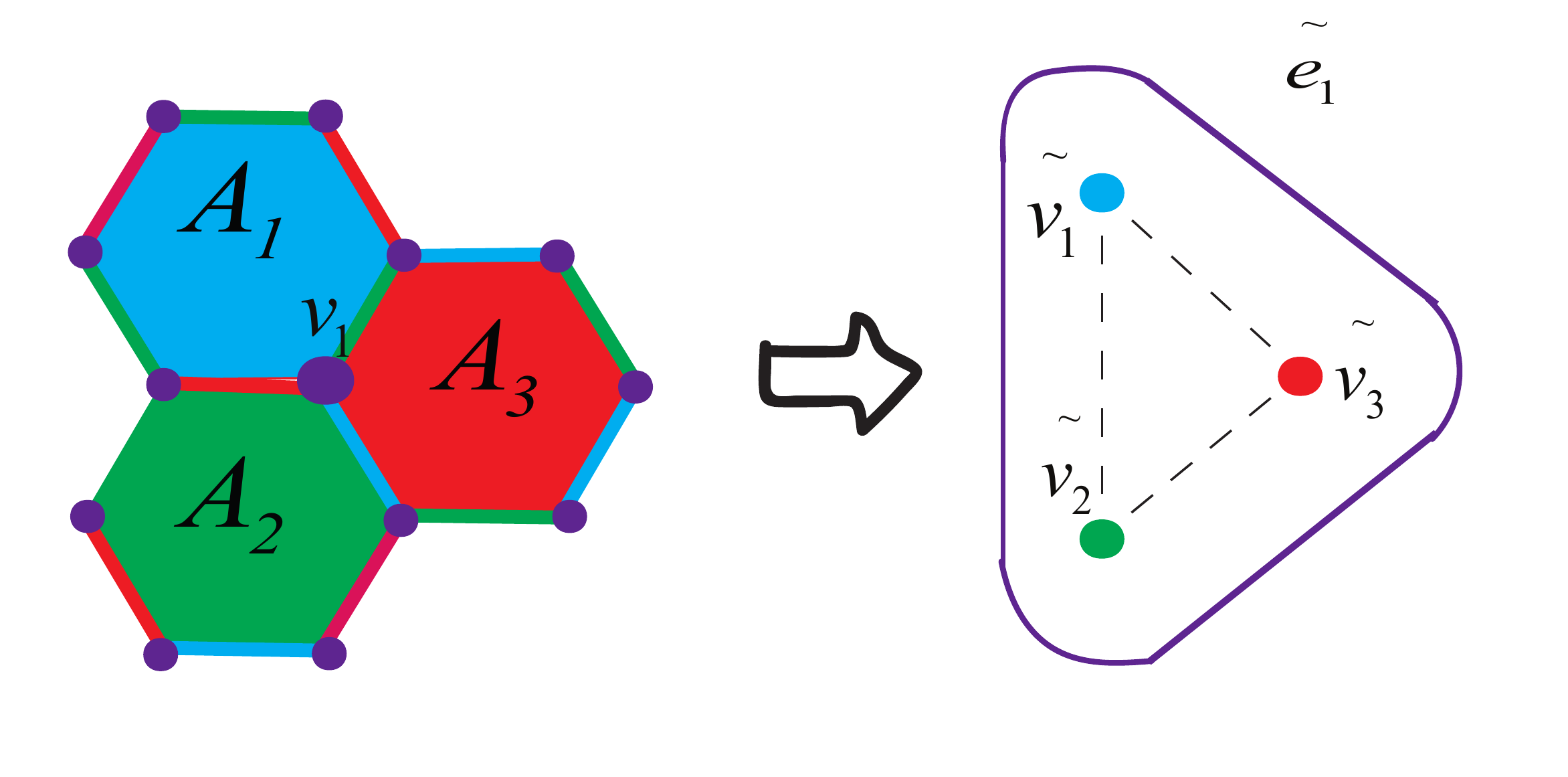}
\caption{(Color online) In a dual space we insert a red vertex corresponding to each hyperedge of the $H$. Since each vertex of the hexagonal lattice is a member of three plaquettes or three hyperedges of the $H$, in a dual space each hyperedge involves three vertices of the $\tilde{H}$ denoted by a triangle.} \label{colohyp}
\end{figure}

\begin{figure}[t]
\centering
\includegraphics[width=8cm,height=4.5cm,angle=0]{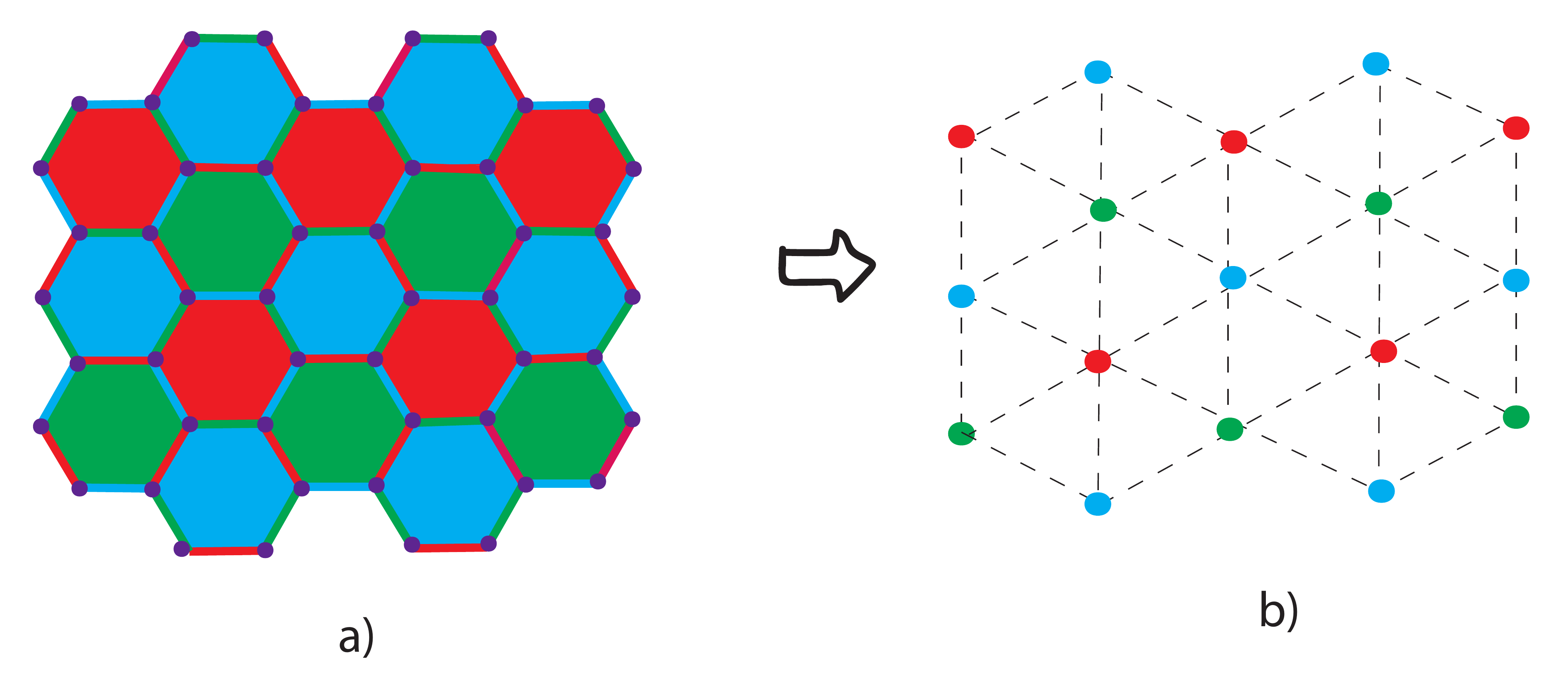}
\caption{(Color online) By applying a transformation similar to
Fig.(\ref{colohyp}) for all vertices, we will have a triangular
lattice in the dual space with a new Hamiltonian with three-body
interactions corresponding to each triangle of the lattice.}
\label{colexa}
\end{figure}

\begin{figure}[t]
\centering
\includegraphics[width=8cm,height=4.5cm,angle=0]{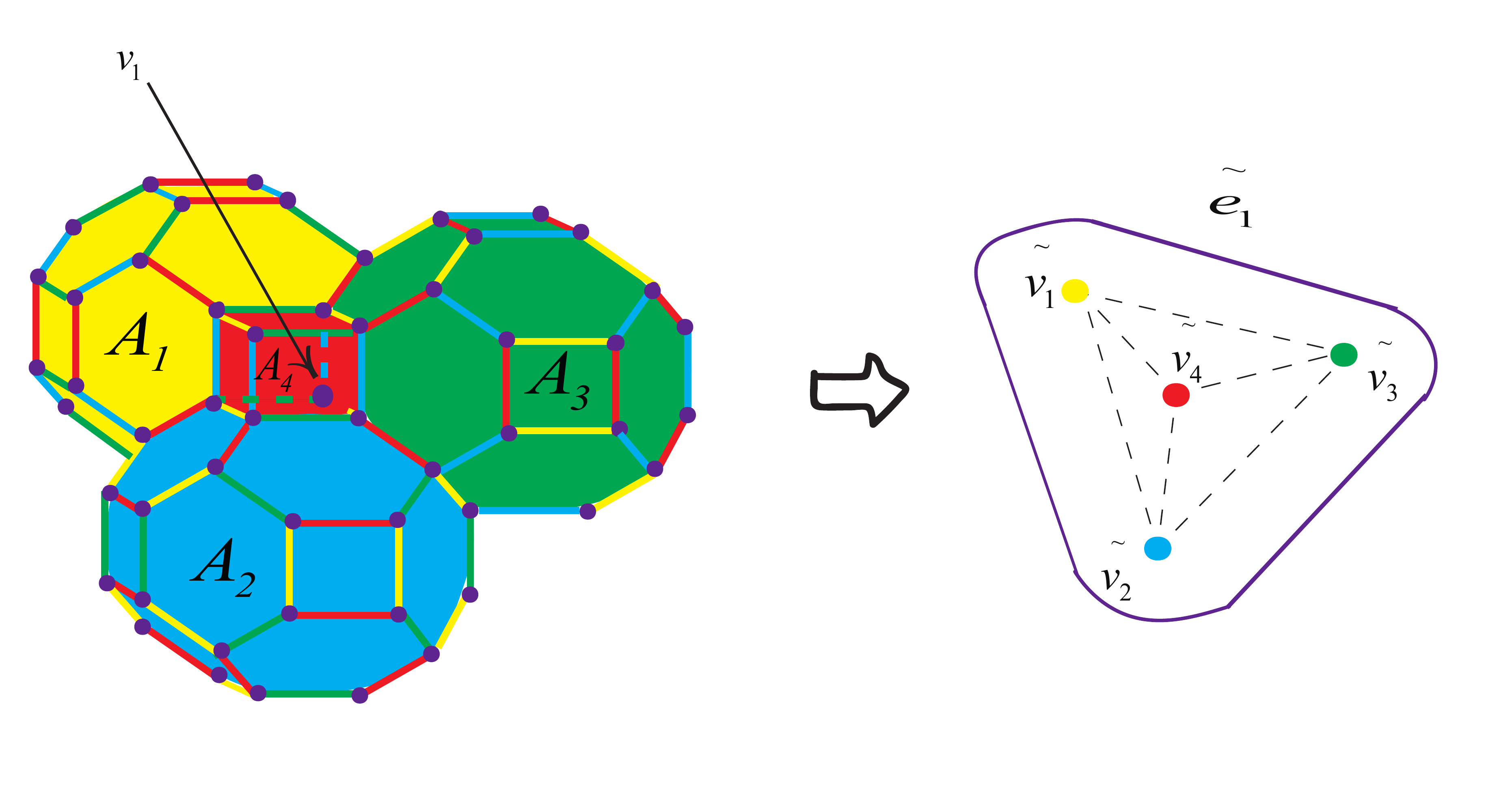}
\caption{(Color online) In a dual space we insert a red vertex corresponding to each hyperedge of the $H$. Since each vertex of a 3-colex is a member of four cells of the lattice or four hyperedges of the $H$, in a dual space each hyperedge involves four vertices of the $\tilde{H}$ where we will have a tetrahedron lattice with a Hamiltonian with four-body interactions corresponding to tetrahedrons.} \label{colo3}
\end{figure}

Now we define a hypergraph $H$ corresponding to this state where
each qubit will be a vertex of the hypergraph and each plaquette
of the 2-colex corresponds to a hyperedge of the $H$ which
involves all qubits belonging to that plaquette. Then we should
find dual of such a hypergraph. As it is shown in
Fig(\ref{colohyp}), since each vertex $v$ of the $H$ is a member
of three neighboring hyperedges, in the dual space each edge
$\tilde{e}$ should involve three vertices. Therefore, dual
hypergraph will be a triangular lattice where triangles
corresponds to hyperedges of the $\tilde{H}$, see
Fig(\ref{colexa}). According to the dual mapping, the spin model
corresponding to the $\tilde{H}$ is a Ising-like model with
three-body interactions in a transverse field.
Then consider a CC on a 3-colex. Since in this case $X$-type operators are related to cells of the 3-colex, a hypergraph $H$ should be defined with hyperedges corresponding to the cells. The next step is to find dual of such a hypergraph. As it is shown in Fig(\ref{colo3}), since each vertex is a member of four hyperedges of the $H$, in the dual space each hyperedge would involve four vertices. In this way, the $\tilde{H}$ is related to a tetrahedron lattice where each tetrahedron corresponds to a hyperedge of the $\tilde{H}$ which involves four vertices belonging to that tetrahedron. According to the dual mapping, the spin model corresponding to the $\tilde{H}$ will be an Ising-like model on a tetrahedron lattice with four-body interactions corresponding to each tetrahedron and in presence of a transverse field.
Extension of the above idea to CC in higher dimensions is straightforward. In fact, it has been shown that dual of a $D$-colex is a D-simplicial lattice with $(D+1)$-colorable vertices \cite{Bombin2015}. Therefore, According to the dual mapping, a CC on a $D$-colex in magnetic field is mapped to a Ising-like model with $(D+1)$-body interactions corresponding to each cell of a $D$-simplicial lattice in presence of a transverse field. Unfortunately, such spin systems on simplicial lattices are very abstract and they have been studied only for some specific two-dimensional examples \cite{baxterwu}. Therefore, unlike TC, we will not be able to compare the robustness of CC on different lattices.

\subsection{Self-dual models}
As we mentioned in section (\ref{s1}), a hypergraph will be called self-dual if it is the same as its dual. On the other hand, according to our duality mapping, for a Hamiltonian on a self-dual hypergraph, the dual Hamiltonian will be the same as the original Hamiltonian where the coupling $J$ and magnetic field $h$ has been exchanged. In this way, it is clear that such a model will have a phase transition at $\frac{h}{J}=1$ if there are two different phases in two limits $h$, or $J$ goes to zero. In the following of this subsection we give some examples of such models.

\begin{figure}[t]
\centering
\includegraphics[width=6cm,height=2.5cm,angle=0]{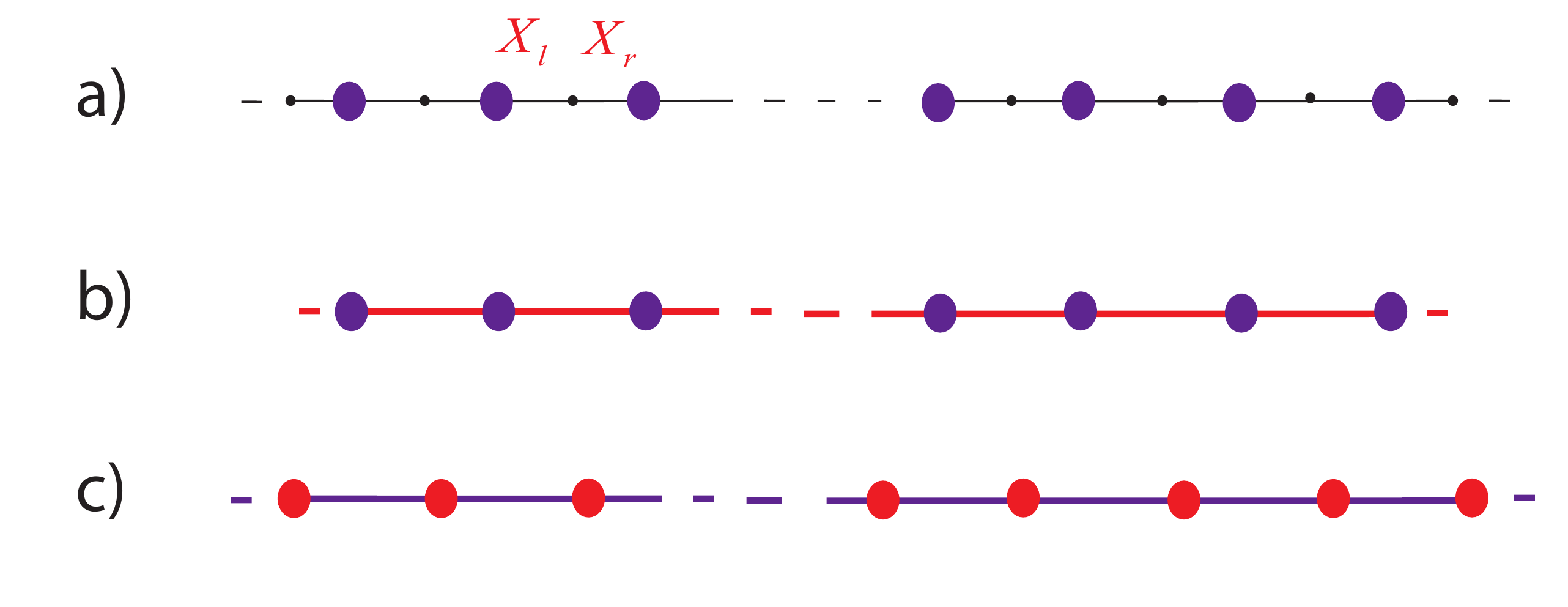}
\caption{(Color online) a) A one-dimensional lattice where qubits
live in edges of the lattice and an $X$-type stabilizer is defined
corresponding to each vertex. b) A hypergraph representation for
the above model where corresponding to each stabilizer we have
defined a hyperedge denoting by red color. c) In a dual space we
insert a red vertex corresponding to each hyperedge of the $H$.
Since each vertex of the $H$ is a member of two hyperedges, in the
dual space each hyperedge also involves two vertices of the
$\tilde{H}$. Therefore, the $\tilde{H}$ will be the same as the
$H$.} \label{one}
\end{figure}
The first example is a one-dimensional model. To this end consider
a one dimensional lattice with qubits which live in edges of the
lattice. Corresponding to each vertex of the lattice we define a
$X$-type operator in the form of $A_v =X_l X_r$ where $l$ and $r$
denote qubits in the left-hand and right-hand of each vertex. In
order to define a hypergraph corresponding to such model it is
enough to relate a hyperedge to each $X$-type stabilizer. In
Fig.(\ref{one}) we show such a hypergraph. The number of vertices
and hyperedges for this hypergraph are the same and since each
vertex is a member of two neighboring hyperedges, it is simple to
check that such a hypergraph is self dual, see Fig.(\ref{one}). In
this way, according to the duality mapping a Hamiltonian
corresponding to the above $X$-type stabilizers in a uniform
magnetic field is self-dual and the phase transition point will be
at $\frac{h}{J}=1$. Such a Hamiltonian is a one-dimensional Ising
model in a transverse field and the above phase transition point
is a well-known result.

\begin{figure}[t]
\centering
\includegraphics[width=8cm,height=2.5cm,angle=0]{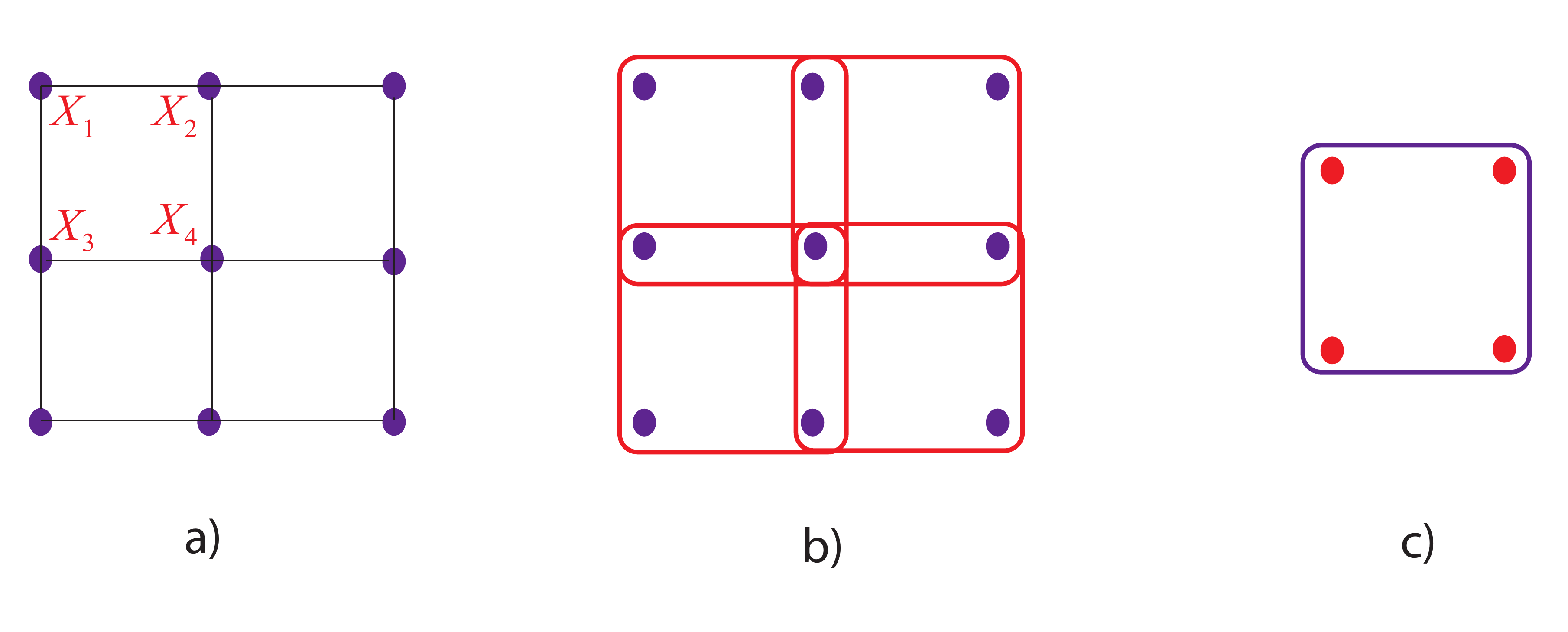}
\caption{(Color online) a) A two-dimensional square lattice where
qubits live in vertices of the lattice and an $X$-type stabilizer
is defined corresponding to each plaquette. b) A hypergraph
representation for the above model where corresponding to each
stabilizer we have defined a hyperedge denoting by red curves. c)
In a dual space we insert a red vertex corresponding to each
hyperedge of the $H$. Since each vertex of the $H$ is a member of
four hyperedges, in the dual space each hyperedge also involves
four vertices of the $\tilde{H}$. Therefore, $\tilde{H}$ is the
same as the $H$.} \label{two}
\end{figure}

Another example of a self dual model can be defined on a two-dimensional square lattice where qubits live in vertices. Corresponding to each plaquette of the lattice we define a $X$-type stabilizer in the form of $A_p =X_1 X_2 X_3 X_4$ where 1, 2, 3, 4 denote qubits on four vertices belonging to each plaquette. In Fig.(\ref{two}), we show a hypergraph representation for such a model. In a square lattice the number of vertices and plaquettes are the same and since each vertex is a member of four hyperedges of the hypergraph, it is simple to check that such a hypergraph is self-dual. In this way the phase transition point of such a model in a transverse field will be at $\frac{h}{J}=1$.

Extension of the above two examples to higher dimensions is straightforward. In three dimension, it is enough to define $X$-type stabilizers corresponding to each cubic cell of the lattice and it will be clear that the corresponding hypergraph will be self-dual. Generally in $D$ dimension we can define a rectangular lattice and corresponding to each D-dimensional unit cell we should define an $X$-type stabilizer. Such a model in a transverse field is also self-dual with a phase transition point at $\frac{h}{J}=1$.
\section{Discussion}
Mapping quantum many-body systems to hypergraphs is an
interesting idea which should be followed in the future. In this
paper, we used such an idea for studying quantum CSS codes in a
uniform magnetic field at zero temperature. Especially, we derived
a strong-weak coupling duality that is specifically interesting in
view of theoretical physics where quantum phase transitions for
two different quantum many-body systems are mapped to each other
in two switched regimes of couplings . Especially, we used such a
mapping for considering one of the important problems in quantum
information theory namely the robust quantum memory. By studying
the robustness of TC on different graphs, we showed that the
robustness of TC defined on graphs decreases in higher dimensions. On the other
hand, it has already been well-known that power of
self-correctness of a topological code in finite temperature
increases in higher dimensions. We should emphasize that we achived to
such a result just for a subclass of TC defined on graphs, with qubits living on edges, which are not useful as self-correcting quantum memores in any dimension due to their string-like excitations. Therefore,  it will be interesting that one
studies other topological codes in this direction. In other words, our results propose
that there might be an optimum dimension where we might have an
efficient quantum memory for a more general topological code. Furthermore, we
used self-duality of hypergraphs to introduce several self-dual
models where we exactly found the phase transition point.

\end{document}